# An Enhanced Multiple Random Access Scheme for Satellite Communications

Huyen-Chi Bui[1,2], Jérôme Lacan[1] and Marie-Laure Boucheret[2]
[1]University of Toulouse, ISAE/DMIA,
[2]University of Toulouse, IRIT/ENSEEIHT
Email: {huyen-chi.bui, jerome.lacan}@isae.fr, marie-laure.boucheret@n7.fr

*Abstract*—In this paper, we introduce Multi-Slots Coded ALOHA (MuSCA) as a multiple random access method for satellite communications. This scheme can be considered as a generalization of the Contention Resolution Diversity Slotted Aloha (CRDSA) mechanism. Instead of transmitting replicas, this system replaces them by several parts of a single word of an error correcting code. It is also different from Coded Slotted ALOHA (CSA) as the assumption of destructive collisions is not adopted. In MuSCA, the entity in charge of the decoding mechanism collects all bursts of the same user (including the interfered slots) before decoding and implements a successive interference cancellation (SIC) process to remove successfully decoded signals. Simulations show that for a frame of 100 slots, the achievable total normalized throughput is greater than 1.25 and 1.4 for a frame of 500 slots, resulting in a gain of 80% and 75% with respect to CRDSA and CSA respectively. This paper is a first analysis of the proposed scheme and opens several perspectives.

## I. INTRODUCTION AND RELATED WORK

Since its apparition 40 years ago, the ALOHA protocol [1] developed at the University of Hawaii and its variants, have motivated an extremely large number of studies. The pure ALOHA is a protocol for sharing channel access among a number of users wishing to send at anytime short data packets with relatively low throughput demand. The optimal normalized throughput $T$ of this scheme is equal to $0.18$. The next versions of ALOHA, the Slotted ALOHA (SA) [2] or its enhanced version named Diversity Slotted ALOHA (DSA) [3] where time is split into slots are used in satellite networks for transmission of short bursts. In SA, the users send packets at fixed time slots of one packet length, and the maximum normalized throughput is doubled compared to the pure Aloha protocol ($0.37$ vs. $0.18$). In DSA, the same packet is transmitted twice to improve the throughput and delay but at low normalized load. This scheme does not provide a great enhancement with respect to SA.

After DSA, improved versions (CRDSA*) named Contention Resolution Diversity Slotted ALOHA (CRDSA) [4] and CRDSA++ [5] have been developed and included into the second generation of Digital Video Broadcasting - Return Channel via Satellite (DVB-RCS2) standard [6]. Similarly to DSA, 2 replicas (3 to 5 replicas in the case of CRDSA++) of the same packet are generated and sent randomly onto the frame of $N_s$ slots. The improvement is that each packet contains a signaling information which points to its replicas location. Whenever one packet is successfully decoded, the replicas are also located and canceled. This procedure is iterative until no decoding is possible.

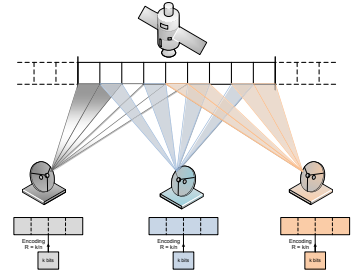

Fig. 1. Multiple access on a slotted channel

Recently, a generalized scheme of CRDSA was introduced: Irregular Repetition Slotted ALOHA (IRSA) [7] which allows a variable repetition rate for each burst to provide a higher throughput gain over CRDSA. Using this protocol in a system where $N_s$ is equal to 200 can achieve a normalized throughput close to $0.8$.

Afterwards, a new generalization of IRSA named Coded Slotted ALOHA (CSA) [8] was also introduced. CSA encodes the bursts using local codes before the transmission. The maximum achievable throughput is $0.8$ when the following assumptions are satisfied: sufficiently high SNR, ideal channel estimation and erasure of bursts in collision.

In this paper, we introduce a further generalization of CRDSA named Multi-Slots Coded ALOHA (MuSCA). We consider a system where several users share a channel to send data to a given access point or relay such as a satellite, a base station (see Figure 1 for an example where a satellite acts as a relay). All users encode their data with an error correcting code to generate one codeword which is cut into $N_b$ physical layer packets ($N_b$ varies from one user to other) of same length called bursts. The bursts are transmitted on a multiple access frame which is split into logical slots. Each user randomly sends its $N_b$ bursts in $N_b$ different time or frequency slots so bursts may collide. The received signal (which can be either at the level of the relay, the gateway or the terminal) is composed by the noisy sum of the signals relative to the different users. The codeword of each user is rebuilt from the $N_b$ bursts including collided ones by the receiver. Bursts collisions are almost cleaned up by successive decoding and interference cancellation operations. The objective of our paper is to generalize the CRDSA* mechanism, to underly the proposed enhancements and to present the significant gain in terms of throughput and packet loss ratio ($PLR$) over different schemes. Subsequently, we show (under our hypothesis) how a peak throughput of 1.3 can be reached by MuSCA.

The paper is organized as follows: the proposed multiple access scheme is presented in the next section. The implementations of our mechanism is detailed in section III. Then, an analysis in terms of throughput and packet loss ratio ($PLR$) according to different system parameters is provided in section IV, allowing performance comparisons between MuSCA and existing schemes presented in section I. Future work and conclusion are given in the end.

## II. SYSTEM OVERVIEW

In this section we first present the assumptions taken on the communication system. Then, we detail the decoding mechanism used in the receiver which combines classical error decoding and successive interference cancellation (SIC).

### A. Hypotheses

We consider the uplink of a satellite communication system shared among $N_u$ users. The access method is based on slotted ALOHA where the communication medium is divided into slots of one burst size. The burst size is the same for all the users. We consider that a set of $N_s$ consecutive slots forms a frame. In this paper, the channel is considered linear and the transmission is subject to an Additive White Gaussian Noise (AWGN).

In our system, a user can only send one data block (i.e., $N_b$ bursts) on one frame. To continue sending more messages, the user must wait until the beginning of the next frame. Bursts of all users are mapped onto slots. Here, we assume that synchronization mechanisms allow the users to be synchronized at slot and frame levels.

If several users attempt a transmission during the same slot, there is a collision. Contrary to CSA, in this work, collided bursts are not considered as erased, and they are integrated into the decoding process.

### B. Principle of the Mechanism

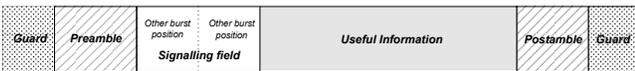

Fig. 2. Burst structure

*1) Transmitter:* In this system, each user wishes to transmit a data block of $k$ bits on each frame. First, an error-correcting code of rate $R$ is used. A good example of such code is the 3GPP2 turbo code, also used in DVB-SH, which is able to simultaneously manage errors and collisions [9]. A block of $1/R \times k$ bits is generated. Encoded data block is then interleaved and split into $N_b$ bursts. Similarly to CRDSA*, signaling information bits and a preamble are added to each burst. The structure of each burst is represented in Figure 2.

Modulated bursts are sent on several slots of a slotted-ALOHA-like channel (see Figure 3). The signaling bits contain information identifying the positions of the other bursts of the same user within the frame. Theses bits are encoded with a short code of rate $R_s$. Given a target $E_b/N_0$ ($E_b$ being the energy per bit and $N_0$ the noise power spectral density), the code rate $R$ is determined in order to guarantee a given Bit Error Rate (BER). We assume that codes of all users can be different and are known by the receiver.

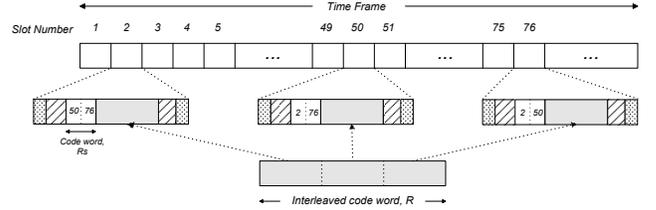

Fig. 3. Sending scheme, time-hopping case, $N_b = 3$

*2) Receiver:* The received signal on a frame is the sum of signals after passing through the channel of the $N_u$ users. The Interference Cancellation (IC) process is applied twice to this signal: first to decode signalization fields and then to decode data fields of each located users' set of $N_b$ bursts.

First, the receiver attempts to locate bursts of as much users as possible by decoding the signaling fields with the Successive Interference Cancellation (SIC) algorithm. It runs through each slot, tries to decode the signalization field with the decoder of rate $R_s$. If the decoding is successful, the positions of the $N_b - 1$ other bursts of the same user are discovered. This allows the receiver to generate the $N_b$ signalization fields and subtract them from the received signal. After this operation, $N_u - 1$ users remain to be located. The process is iterative until the end.

Afterwards, data decoding is started for the located user who has the most non collided (clean) bursts. The receiver collects all the bursts of the same user, rebuilds a code word of length $1/R \times k$ and sends it to the decoder. In CSA, collided bursts are treated as erasure. In MuSCA, when a burst is located, even if it is interfered by bursts of other users, it can still participate in the decoding process. A burst is considered as lost only when it is highly interfered and brings no more useful information for the decoding. The threshold for high interference is defined according to the code used by each user. The integration of collided bursts into the decoding process significantly improves the system performance in terms of throughput. If the decoding is successful, the receiver creates the $N_b$ bursts of this user. After that, the signals corresponding to the $N_b$ recovered bursts are subtracted with interference cancellation processing from the received signal. The interference contribution caused by these bursts is so removed. After the subtraction, the resulting signal is the combination of channel noise and signals of the $N_u - 1$ remaining users. The decoding algorithm is iterative until the occurrence to a deadlock situation where no user is still decodable.

Figure 4 shows a case of deadlock situation for CSA and its previous versions (CRDSA, IRSA). In this case, with a threshold set to 2, MuSCA is able to implement the decoding even if there is no more clean burst on the frame.

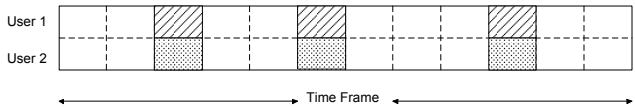

Fig. 4.  Decodable by MuSCA but not by CRDSA*, IRSA and CSA

## III. IMPLEMENTATION

The algorithm described in the previous section allows the system to achieve a high throughput. However, the choice of codes for the signalization field, payload and the parameters $N_b$, $N_s$ plays an important role that may deeply influences the system performance.

Although the conception of the algorithm allows the use of a different code for each user (which could be beneficial for the system performance), we only detail the case where users encode their data with the same error correcting code of rate $R$ and then the same $N_b$. Several studies on CRDSA and CSA [5] [8] show that these systems achieve their best performance in terms of throughput for $N_b = 3$ (normalized throughput $T = 0.7$ for CRDSA++ and $T = 0.8$ for CSA). Therefore, in the following sections, only results with the parameter $N_b$ equal to 3 are presented.

### A. Signalization fields

As in CRDSA, the signaling field of each burst contains pointers to the positions where the other bursts of the same user are located. The field size depends on the size $N_s$ of the frame but remains relatively small. Signaling data are encoded, modulated (e.g., using BPSK modulation) and added to the useful data field to form a burst. Reed-Muller codes are short codes which achieve an excellent trade-off performance/complexity for soft-decision decoding [10]. Therefore, they can be used to encode these signaling fields. The signalization decoding is launched in the case that the burst is on a collision-free slot or it is interfered by only one other user.

### B. Data field

The algorithm proposed by MuSCA does not require any constraint on the choice of codes for payload data block. However, in order to compare with the existing methods, we have first conducted studies with a code allowing to obtain the same length of burst and to send an equivalent quantity of data per slot as CRDSA. In the CRDSA* scheme, a terminal sends $N_b$ copies of the same MAC packets in 3 randomly selected slots, the payload in each burst is encoded by a convolutional [4] or turbo code [5] of rate $r = \frac{1}{2}$. This is equivalent to a general code of rate $R = \frac{1}{2N_b}$. In MuSCA, a CCSDS turbo encoder of rate $R = 1/6$, associating with QPSK modulation can be applied to bit sequences of length $k = 456$, producing coded blocks of 2760 bits, equivalent to 1380 symbols. These blocks are then randomly interleaved, split into 3 bursts of length about 460 symbols and sent on slots. Figure 5 depicts the performance curves of this turbo code in the 3 cases: a) the 3 bursts are on clean slots, b) the 3 bursts are in collision with one user's signal and c) the 3 bursts are in collision with 2 other users. Other analysis using a different code will be carried out in order to increase the throughput reached by the system (for example, a turbo code with $R = 1/4$ and length $k = 680$ bits). Note that in our work, the system performances are shown with simple CCSDS codes. It is possible to improve them by using specially designed codes for this context.

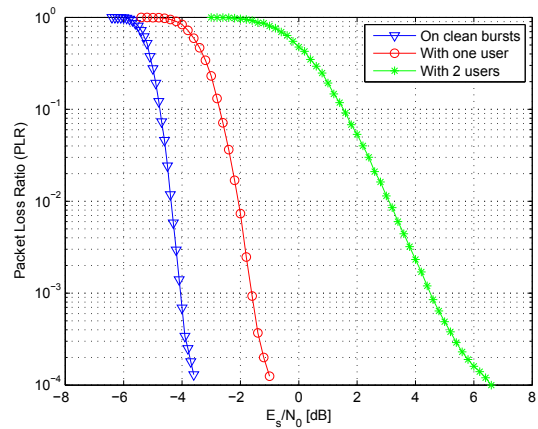

Fig. 5.  Turbo code of R = 1/6, k = 456 bits, modulation QPSK, AWGN channel

## IV. PERFORMANCE EVALUATION

Our analyses rely on the assumptions that users arrive on the communication medium in a perfectly synchronized manner (both in timing and phase) and under the constraint of equal average received power. The MuSCA decoding algorithm using SIC attempts to remove the interference of the the most recently decoded user, by re-encoding the decoded bit sequence, modulating it with the appropriate amplitude and phase adjustment, and subtracting it out from the current composite received signal. To realize this operation, we assume that the channel estimation is perfect.

We still consider a frame composed of $N_s$ slots, in which $N_u$ users attempt a transmission. The normalized load representing the average number of blocks transmissions per slot is computed as:

$$G = \frac{N_u}{N_s} \quad (1)$$

The probability of non decoding a packet is denoted $PLR$ and is given for each $E_s/N_0$ and $G$. For a fixed $E_s/N_0$, the relation between normalized throughput (defined as the probability of successful block transmission per slot) and $PLR$ is given by:

$$T = G \times (1 - PLR(G)) \quad (2)$$

The value of $G$ that maximizes $T$ must be carefully chosen. Indeed, $T$ is bounded by $G$ so with a greater $G$, the expected normalized throughput can be higher. But if $G$ exceeds a

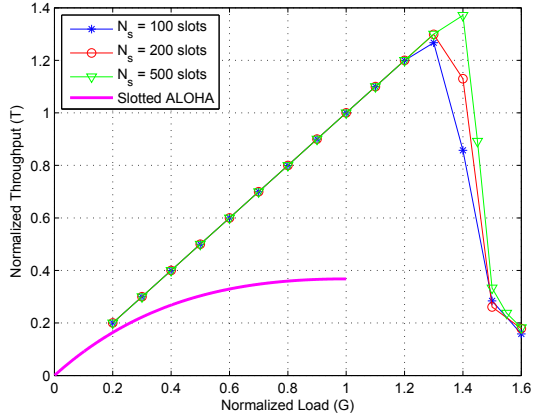

Fig. 6. Normalized throughput for MuSCA using turbo code R=1/6; various frame sizes

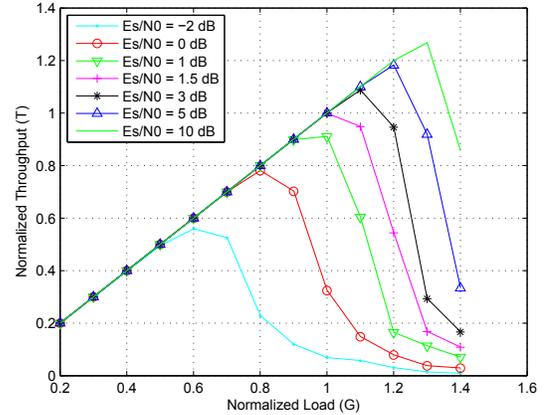

(a) MuSCA

certain value, the frame is full, the collision rate is high, which causes a large packet loss ratio and reduces $T$.

In the first set of simulations, a turbo code of rate $R = 1/6$ is used. On one hand, we consider that bursts interfered by 3 or more other users are not reliable. They are considered as erased (the threshold is set to 2). On the other hand, taking into account bursts which are interfered by only one or two other users can significantly ease the decoding and thus increases the throughput.

In Figure 6, the throughput is evaluated for different frame sizes $N_s = 100, 200$ and $500$ when $E_s/N_0 = 10dB$. The curve reached by Slotted ALOHA $T = G \times e^{-G}$ is also presented for reference. The gain between MuSCA and other methods is significant. For the frame length $N_s = 1000$, CSA approaches $T \approx 0.8$ with the noiseless assumption [8], while MuSCA achieves a peak throughput of about 1.4 for $E_s/N_0 = 10dB$ with a half frame size.

We can observe that the frame size does not have a significant impact on the performance of MuSCA: 1.4 for $N_s = 500$ vs. 1.25 for $N_s = 100$. In the same time, a large augmentation of $N_s$ would increase significantly the transmission delay because the decoding process cannot be started before the entire frame is received. For example, the utilization of frame formed by 500 slots would introduce a delay 5 times longer. For this reason, a frame size of 100 slots has been assumed for all following simulations.

The system performance in terms of throughput is then studied by varying the signal to noise ratio ($SNR$) and finding the normalized load $G$ that maximizes the normalized throughput $T$ for each $SNR$. A frame is split into 100 slots. From Figure 7, we see that MuSCA is able to achieve significant better thresholds than CDRSA for every $E_s/N_0$. On a frame of this length, the system can transmit up to 125 users ($T = 1.25$) for a high $SNR$ ($E_s/N_0 = 8dB$) while CRDSA is limited to less than 70 users. MuSCA with turbo code $R = 1/6$ provides a peak of throughput close to 1.3 and the relation between $T$ and $G$ is almost linear up to $G$ close to 1.2. That means even if the communication medium is 120% loaded, the probability of successful transmission is almost 100%. When the power of noise is equal to the useful signal ($E_s/N_0 = 0dB$), MuSCA still allows up to 80 users to transmit, whereas the CRDSA system is completely overridden.

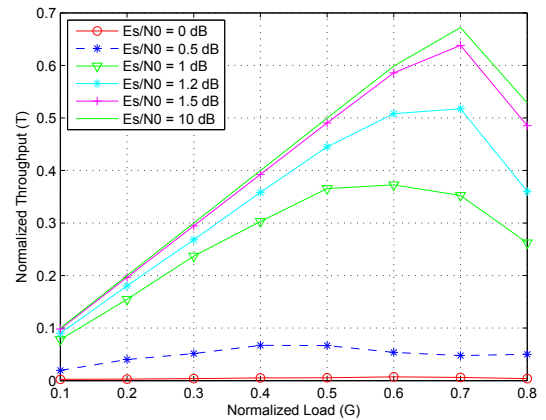

(b) CRDSA, 3 replicas

Fig. 7. Throughput $T$ versus the normalized load $G$ for MuSCA and CRDSA. $N_s = 100$ slots

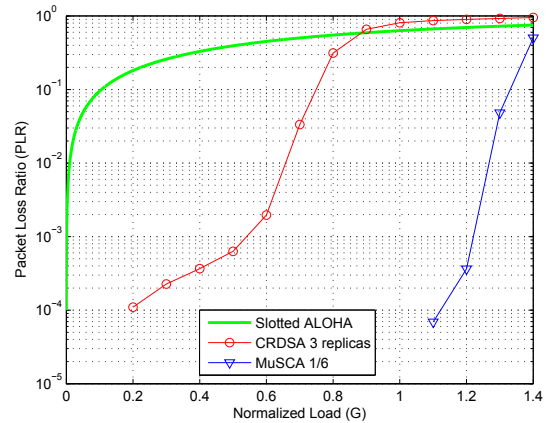

Fig. 8. Packet Loss Ratio for SA, CRDSA with 3 repetitions and for MuSCA. $E_s/N_0 = 8dB$

Figure 8 shows the $PLR$ as a function of $G$ for SA, CRDSA

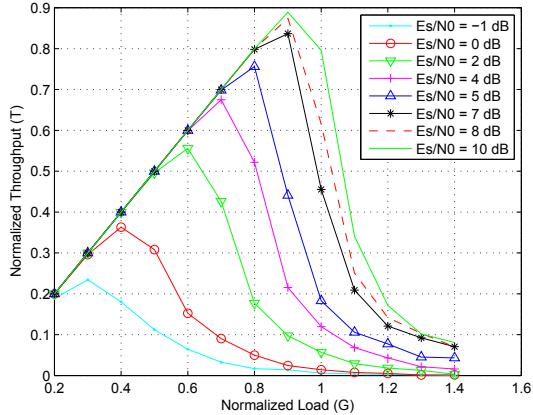

Fig. 9. Throughput $T$ versus the normalized load $G$ for MuSCA using turbo code 1/4, $N_b = 3$ and $N_s = 100$ slots

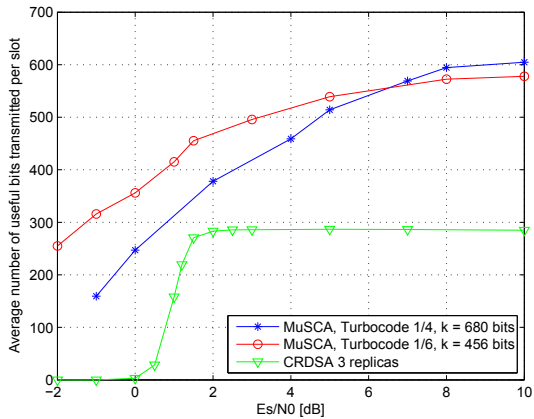

Fig. 10. Quantity of useful bits transmitted per slot for MuSCA, CRDSA 3 replicas vs. turbo code 1/6 and turbo code 1/4, $N_s = 100$ slots

and MuSCA. MuSCA can achieve a remarkable low $PLR$ (less than $0.1\%$) at a high normalized load $G = 1.2$. Consequently, most of bursts are successfully transmitted at the first attempts, resulting in low latencies. In this figure, we observe that MuSCA has much lower PLR for all loads in the case of no re-transmission. For a $PLR$ equal to $10^{-3}$, the MuSCA channel can be loaded 2.3 times more than the CRDSA channel ($G_{MuSCA} \approx 1.22$ vs. $G_{CRDSA 3 replicas} \approx 0.54$) and 1220 times more than the SA channel ($G_{SA} \approx 0.001$). Targeting a $PLR$ equal to $10^{-2}$, two and three replicas CRDSA can offer a traffic close to 0.35 [5] and 0.66 respectively. SA only operates at extremely low load $G_{SA} \approx 0.01$ while for MuSCA, this $PLR$ can be achieved with a channel traffic $G \approx 1.27$.

The same simulations are executed with the second turbo code. We consider a code of rate $R = 1/4$ to encode sequences of 680 bits, combined with QPSK modulation. Then 1380 QPSK symbols codewords are obtained. As previously, $N_b$ is set to 3. That means each user tries to transmit about 1.5 times more useful information on each frame and on each slot. The throughput of the MuSCA protocol has been simulated versus the normalized load $G$ for variable values of $E_s/N_0$.

Performance results are given in Figure 9.

By combining numerical results from both simulations sets, we quantify the number of maximum payload bits for each code. Focusing on the relationship between the number of useful bits transmitted per slot and the $SNR$, we observe that even if the normalized throughput of turbo code of rate $R = 1/4$ is lower than the one of turbo code $R = 1/6$ (peak throughput of 0.9 vs. 1.25), the system still achieves more benefit in terms of useful information sent for high values of $E_s/N_0$. This is due to the low amount of redundancy. In Figure 10, curves of MuSCA using turbo code 1/6 and 1/4 are compared with the curve of CRDSA. All systems operate on frames composed of 100 slots of similar sizes. We note that for any $E_s/N_0$, MuSCA achieves a gain of at least $50\%$ compared to CRDSA.

## V. CONCLUSION AND FUTURE WORK

In this paper, the enhancement so-called Multi-Slots Coded ALOHA (MuSCA) scheme has been introduced and analyzed. The proposed approach represents an improvement of CRDSA, IRSA and CSA protocols. Simulation results show that MuSCA provides a large gain in terms of throughput and packet loss ratio compared to existing random access techniques currently introduced in [4], [5], [7] and [8]. It has been shown that MuSCA can achieve a normalized throughput close to 1.25 for system with frames of 100 slots and 1.4 while the frame size is up to 500 slots for $E_s/N_0$ equal to $10dB$. This results in a 2 and 1.75 fold throughput increase compared to CRDSA and CSA. A low $PLR$ ($10^{-2}$) is also maintained even when the transmission medium is highly loaded ($G$ up to 1.3). In a future work, we expect to further study asynchronous conditions, channel estimation and impacts of estimation errors on the system.